\documentclass[runningheads]{svmult}

\usepackage{makeidx}   
\usepackage{graphicx}  
\usepackage{subeqnar}  
\usepackage{multicol}  
\usepackage{physprbb}  
\makeindex             

\begin{document}
\title*{The masses of AGN host galaxies \& the origin of radio loudness}
\toctitle{The masses of AGN host galaxies
\protect\newline \& the origin of radio loudness}
%
%
\titlerunning{The masses of AGN host galaxies}
%
\author{James S. Dunlop\inst{1}
\and Ross J. McLure\inst{2}}
\authorrunning{James S. Dunlop \& Ross J. McLure}
%
%
\institute{Institute for Astronomy, Royal Observatory, Edinburgh, EH9 3HJ, UK
\and Astrophysics, Department of Physics, Keble Road, Oxford, OX1 3RH, UK}
\maketitle              

\begin{abstract}
We highlight some of the principal 
results from our recent Hubble Space Telescope
studies of quasars and radio galaxies. The hosts of these 
powerful AGN are normal massive ellipticals which lie on the region of the 
fundamental plane populated predominantly 
by massive ellipticals with boxy isophotes and 
distinct cores. The hosts of the radio-loud sources are on average $\simeq 
1.5$ times brighter than their radio-quiet counterparts and appear to lie 
above a mass threshold $M_{sph} > 4 \times 10^{11} {\rm M_{\odot}}$. 
This suggests that 
black holes more massive than $M_{bh} > 5 \times 10^8 {\rm M_{\odot}}$ 
are required to 
produce a powerful radio source. However we show that this apparent threshold 
appears to be 
a consequence of an upper bound on radio output which is a strong 
function of black-hole mass, $L_{5GHz} \propto M_{bh}^{2.5}$. This steep mass 
dependence can explain why the hosts of the most powerful radio sources are 
good standard candles. Such objects were certainly fully assembled by 
$z \simeq 1$, and appear to have formed the bulk of their stars prior to 
$z \simeq 3$.
\end{abstract}

\section{Introduction}

Since the optical identification of Cygnus A (Baade \& Minkowski
1954) it has been clear 
that the host galaxies of the most powerful 
radio sources in the nearby universe appear to be massive ellipticals.
However, it is only in the last decade, since the repair of HST,
that it has proved possible to perform a detailed 
comparison of the hosts of powerful radio-loud and radio-quiet
AGN.
In this brief article we summarize some of the main results from our recent 
HST studies of AGN hosts, with special emphasis on how their structures,
sizes, luminosities and masses compare to those of `normal' galaxies.
We also explore what such studies can teach us about the physical difference
between radio-loud and radio-quiet AGN, and about the formation history
of massive elliptical galaxies.

\section{Quasar host galaxies and `normal' ellipticals}

\begin{figure}
\begin{center}
\includegraphics[width=0.7\textwidth]{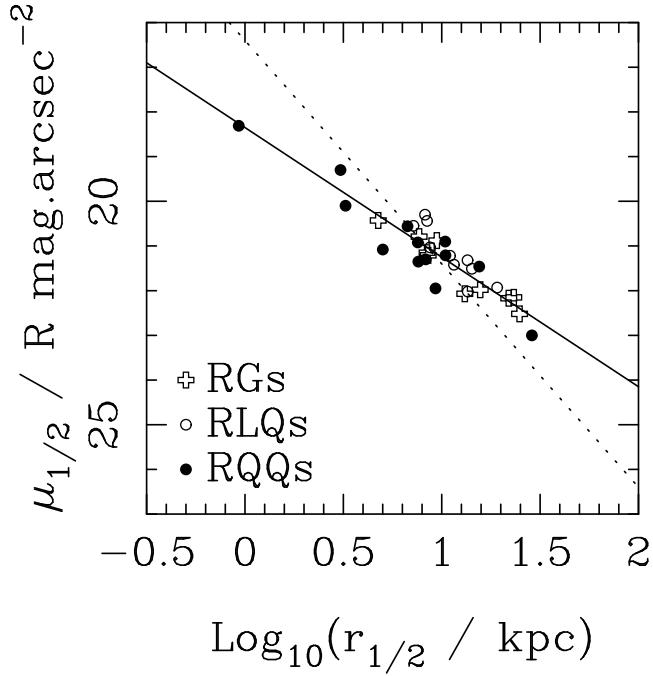}
\end{center}
\caption[]{The Kormendy relation followed by the hosts of all 33 powerful
AGN imaged with the HST by Dunlop et al. (2002).
The solid line is the least-squares fit to the data
which has a slope of 2.90, in excellent agreement with the slope of
2.95 found by Kormendy (1977) for inactive ellipticals in the
$B$-band. The dotted line has a slope of 5, indicative of 
what would be expected if 
the scale-lengths of the host galaxies had not been properly constrained.}
\label{korm}
\end{figure}

One of the most important new results from this work is the discovery not
only that the hosts of powerful AGN (both radio-loud and radio-quiet) 
are almost exclusively ellipticals, but that these galaxies 
display a Kormendy relation 
indistinguishable in both slope and normalization from that displayed by normal
massive ellipticals (Fig. 1; Dunlop et al. 2002). 
The Kormendy relation is the photometric 
projection of the Fundamental Plane (Djorgovski \& Davis 1987; Dressler et al. 1987), but in the case of radio galaxies 
the third dimension (i.e. central stellar velocity dispersion) can be added
with relative ease. This has recently been completed for 
a subset of 22 radio galaxies by Bettoni et al. (2002), who confirm that 
these objects lie towards the bright end of the
same fundamental plane as defined by quiescent massive ellipticals.

The quasar hosts and radio galaxies are therefore all clearly large 
luminous galaxies with $L > L^{\star}$. However, the radio-loud 
hosts are more cleanly confined to a definite high mass regime, with 18/20 of
the radio-loud hosts in the Dunlop et al. sample
having spheroid masses $> 4 \times 10^{11} {\rm M_{\odot}}$, 
compared with only 4/13 of the radio quiet hosts. 
The results of McLure \& Dunlop (2002)
demonstrate that this difference can reasonably be extrapolated to a 
difference in central black holes masses, with the radio-loud sources being
confined to black-hole masses $M_{bh} > 5 \times 10^{8} {\rm M_{\odot}}$.

\section{Radio-power and spheroid/black-hole mass}

At first sight, these results suggest the
existence of a physical mass threshold 
above which galaxies (or their central black holes) are capable of producing 
powerful relativistic jets. 
This would also appear consistent with long-standing 
suggestions of a definite gap in the radio luminosity function of optically
selected quasars. However, the recent study of Lacy et al. (2001) does 
not support the existence of any such gap or threshold. 
In fact Lacy et al. demonstrate 
the existence of a clear, albeit loose, correlation between radio power and 
black-hole/spheroid mass extending over 5 decades in radio power. 
However, the large scatter in the data, and the relatively gentle slope
of the best-fitting relation ($L_{5GHz} \propto M_{bh}^{1.4}$) 
do not provide an obvious explanation of why the hosts 
of powerful radio sources should be such good standard candles.

Instead, Dunlop et al. (2002) have suggested that the distribution of AGN 
on the $L_{5GHz}$:$M_{bh}$ plane is better described as being bounded by a 
lower and upper threshold for the radio output that can be produced by a 
black hole of given mass, and that these radio output thresholds are a 
much steeper function of mass, i.e. $L_{5GHz} \propto M_{bh}^{2.5}$.
In Fig. 2 we demonstrate that the bounding relations deduced by Dunlop et 
al. also provide an excellent description of the data gathered by
Lacy et al.. In fact, the lower boundary is essentially identical to the 
relation derived for nearby galaxies by Franceschini et al. (1998),
who also concluded in favour of $L_{5GHz} \propto M_{bh}^{2.5}$. However,
Fig. 2 makes the interesting (and perhaps surprising) point that 
the upper limit on black-hole radio output appears to be a similarly steep 
function of mass, simply offset by 5 decades in radio power.

This steep upper boundary on $L_{5GHz}$ as a function of 
black-hole/spheroid mass provides a natural explanation for why the 
low-redshift
radio-loud AGN hosts studied by Dunlop et al. lie above an apparently
clean mass threshold. These objects have $L_{5GHz} > 10^{24}
{\rm WHz^{-1}sr^{-1}}$, and from Fig. 2 it can be seen that such radio
powers can only be achieved by black holes with 
$M_{bh} > 2 \times 10^8 {\rm M_{\odot}}$, and hence host spheroids with 
$M_{sph} > 2 \times 10^{11} {\rm M_{\odot}}$. 

Fig. 2 also provides a possible 
explanation for why the 3CR radio galaxies at $z \simeq 1$ appear to be even
better standard candles that at low redshift; inclusion in the 3CR catalogue
at $z \simeq 1$ requires $L_{5GHz} > 10^{26}
{\rm WHz^{-1}sr^{-1}}$, which Figure 3 indicates requires  black holes with 
$M_{bh} > 10^9 {\rm M_{\odot}}$, and hence host spheroids with 
 $M_{sph} > 10^{12} {\rm M_{\odot}}$. At such high masses the luminosity/mass
function of elliptical galaxies is very steep (Kochanek et al. 2001), and 
so it is inevitable that any ellipticals which lie above this mass threshold
will also lie very close to it.

\begin{figure}
\begin{center}
\includegraphics[width=0.7\textwidth]{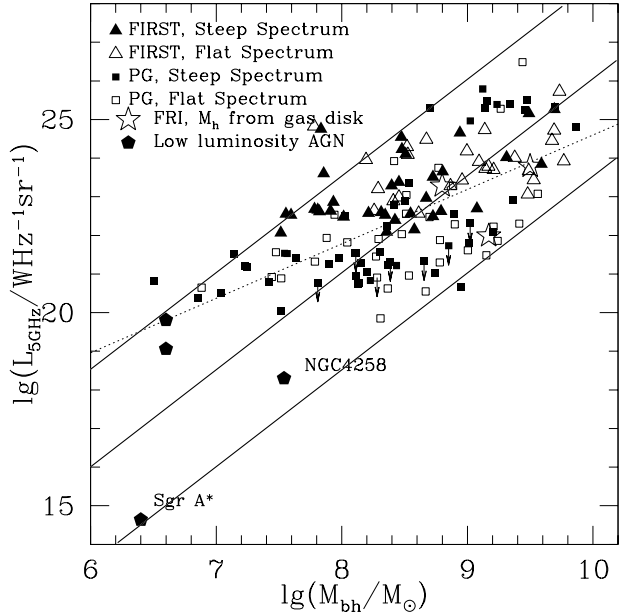}
\end{center}
\caption[]{A plot of lg$L_{5Ghz}$ versus lg$M_{bh}$ compiled from various
samples of quasars by Lacy et al. (2001), with the bounding relations 
on minimum and maximum radio output suggested by Dunlop et al. (2002) 
superimposed (solid lines, $L_{5GHz} \propto M_{bh}^{2.5}$).}
\label{lacy}
\end{figure}

\section{The origin of radio loudness}
At the other end of the radio-power scale, Fig. 2 demonstrates
the surely significant fact that many of the powerful optically-selected AGN
produce a level of radio output which is indistinguishable from that produced
by nearby quiescent ellipticals of comparable mass. In other words, the 
minimum radio power relation defined by the most radio-quiet quasars is 
the same as that defined by nearby `quiescent' galaxies. This dramatically 
illustrates how very different the physical mechanisms for the production 
of optical and radio emission by a black hole must be, since the AGN 
are clearly in receipt of plenty of fuel.

These results therefore lead us to conclude 
that the difference between radio-loud and 
radio-quiet AGN cannot be explained as due to black-hole or host-galaxy mass,
host-galaxy morphology, or indeed black-hole fueling rate. Rather there must
be some other property of the central engine which determines whether
a given object lies nearer to the upper or lower radio-power thresholds
shown in Fig. 2. The only obvious remaining candidate is spin.
This has been previously suggested and explored
by many authors on the basis that angular
momentum must surely be important for the definition of jet direction
(e.g. Wilson \& Colbert 1995; Blandford 2000). Here we have effectively 
arrived at the same conclusion by a process of elimination of the obvious
alternatives.

\section{The assembly of quasar host galaxies}

\begin{figure}
\begin{center}
\includegraphics[width=0.7\textwidth]{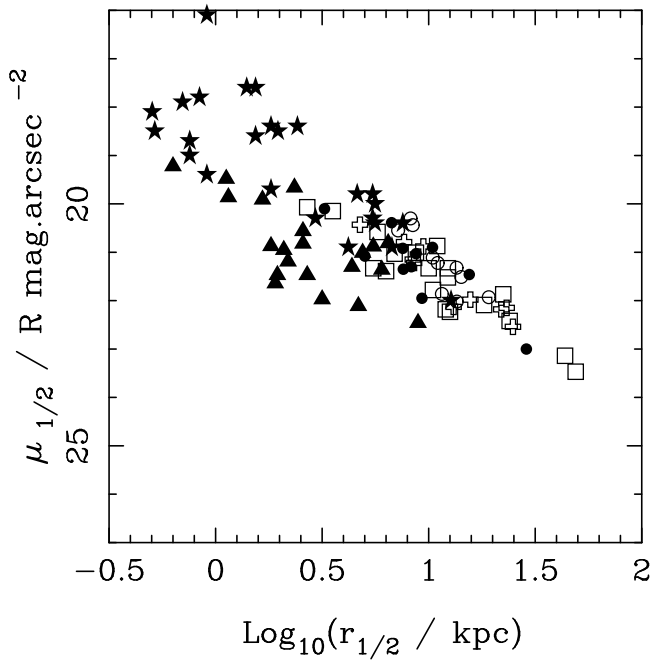}
\end{center}
\caption[]{A comparison of the properties of the AGN hosts with those 
displayed by various other types of spheroid on the photometric 
projection of the fundamental plane. Symbols for the quasar hosts 
and radio galaxies are as in Fig. 1. The stars are the data 
for ULIRGs and LIRGs from Genzel et al. (2001) transformed from the infrared
to the $R$-band assuming $R-K$=2.5. 
Triangles and squares indicate the positions
of `discy' and `boxy' ellipticals from Faber et al. (1997) after conversion 
to $H_0 = 50~{\rm km s^{-1} Mpc^{-1}}$.}
\label{genz}
\end{figure}

A number of independent lines of evidence suggest that the hosts of powerful
AGN formed at high redshift. This evidence is most convincing for the 
radio-loud population: allowing for the effects of passive evolution, radio
galaxies at $z \simeq 1$ lie on the same Kormendy
relation as shown in Fig. 1 (McLure \& Dunlop 2000; Waddington et al. 2002),
the $K-z$ relation for powerful radio galaxies appears consistent with 
purely passive evolution out to $z > 3$ (van Breugel et al. 1998; 
Jarvis et al. 2002), and strong star-formation activity in powerful radio 
galaxies seems largely confined to $z > 2.5$ (Archibald et al. 2001).

The picture is currently somewhat less clear for the hosts of 
radio-quiet quasars. The colours and off-nuclear spectra 
of low-redshift quasar hosts 
indicate that their stellar populations are predominantly old 
(Dunlop et al. 2002; Nolan et al. 2000; McLure et al. 1999) but there is also
some evidence that the hosts of radio-quiet quasars are significantly less
massive by $z \simeq 2$ compared to the present day (Kukula et al. 2001;
Ridgway et al. 2001). This raises the possibility that some of the low-redshift
radio-quiet quasar population could be produced by the same sort of recent
major mergers which power Ultra Luminous Infrared Galaxies (ULIRGS). 
In fact we can 
now begin to explore this possibility directly by combining our own results
of quasar hosts with the results of near-infrared imaging and spectroscopy
recently performed by Genzel et al. (2001).

While it is true that ULIRGs such as Arp220 have surface brightness profiles 
well-described by an $r^{1/4}$-law, Genzel et al. have shown that such 
remnants lie in a different region of the fundamental plane than that which
we have found to be occupied by the quasar hosts. Specifically, the effective 
radii of the ULIRGs is typically an order of magnitude smaller than those
of the quasar hosts. Indeed one can go further and conclude that whereas
ULIRGs may well be the progenitors of the population of 
intermediate-mass ellipticals which display compact cores and cusps
(Faber et al. 1997), the quasar hosts lie in a region of the $\mu_e - r_e$
plane which is occupied by boxy, giant, ellipticals with large cores.
This comparison is illustrated in Fig. 3, where we have augmented the Kormendy
diagram shown in Fig. 1 with the addition of the data from Genzel et al.
on LIRGs and ULIRGs, and the data from Faber et al. (1997) on `discy' and 
`boxy' ellipticals. 
Thus, present evidence suggests that, at least at low redshift, any ULIRG
$\rightarrow$ quasar evolutionary sequence can only apply to a fairly
small subset of objects. 

With the advent of the Advanced Camera on HST, and 
high-resolution near-infrared imaging on ground-based 8-m telescopes, the 
next few years should see some major advances in our understanding of the 
properties of quasar host galaxies as a function of redshift. Fig. 3 indicates
that such studies should also shed light on the formation history 
of the high-mass end of the present-day quiescent elliptical galaxy population.

\end{document}